\documentstyle[preprint,revtex]{aps}                  %%%To be commented
\def\ltap{\ \raisebox{-.5ex}{\rlap{$\sim$}} \raisebox{.4ex}{$<$}\ }
\def\gtap{\ \raisebox{-.5ex}{\rlap{$\sim$}} \raisebox{.4ex}{$>$}\ }

\newcommand{\be}{\begin{equation}}
\newcommand{\ee}{\end{equation}}
\newcommand{\beq}{\begin{eqnarray}}
\newcommand{\eeq}{\end{eqnarray}}

\begin{document}
\draft
\pagestyle{empty}                                      %%%To be commented
\centerline{%Version:\vday
                             \hfill   NTUTH--94--01}   %%%To be commented
\centerline{\hfill                 January 1994} %%%To be commented
\vfill
\begin{title}
Radiative Lepton Masses with Heavy Lepton Seed
\end{title}
\vfill
\author{Gwo-Guang Wong and Wei-Shu Hou}
\begin{instit}
Department of Physics, National Taiwan University,
Taipei, Taiwan 10764, R.O.C.
\end{instit}
%\receipt{\today}
%
%\vskip -1cm
\vfill
\begin{abstract}

We construct a $Z_8$ model for leptons where all Yukawa couplings
are of order unity, but known lepton masses are generated radiatively,
{\it order by order}.
The seed is provided by fourth generation leptons $E$ and $N$,
which receive (Dirac) mass in usual way.
Two additional Higgs doublets with nontrivial $Z_8$ charge
are introduced to give nearest neighbor Yukawa couplings.
Hence, nonstandard Higgs bosons are flavor changing in an unusual way.
Loop masses are generated when $Z_8$ is {\it softly} broken down to $Z_2$.
However, $e$ and $\mu$ mass generation require new Higgs bosons to be
at weak scale. Neutral scalar mixing underlies
$m_\mu/m_\tau \gg m_e/m_\mu$, $m_\tau/m_E$.
The $Z_2$ symmetry forbids $\mu\to e\gamma$ and $\tau\to \mu\gamma$.
The most stringent bound comes from $\tau\to \mu\mu^\pm e^\mp$.
The model has interesting implications for $\tau\to e\gamma$,
$\mu \bar e \to \bar\mu e$ conversion, $\mu\to e\nu_e\bar\nu_\mu$,
and FCNC decays of $E$ and $N$ (such as $E\to \tau e^\pm\mu^\mp$).

\end{abstract}
%%%%%%%%%%%%%%%%%%%%%%%%%% See PRL 69 #24. 1992 for the listings
%\newpage
%
\narrowtext
\pagestyle{plain}

%\section{Introduction}

A major mystery regarding fermion flavor is the very wide range
of its mass spectrum. Neutrinos appear to be massless,
while known masses range from the electron's
$0.511$ MeV, to over 120 GeV \cite{TeVnew} for the top quark.
In the standard $SU(2)_L \times U(1)$ electroweak model (SM),
a single Higgs doublet is responsible for generating all masses.
The natural scale is $v=( \sqrt 2 G_F)^{-1/2} \approx 246$ GeV,
where $G_F$ is the Fermi coupling.
The puzzle is then two fold.
On one hand, the dimensionless Yukawa coupings $f$
are scattered over a wide range,
and there is an empirical family hierarchy, {\it e.g.}
$f_e \ll f_\mu \ll f_\tau$, for each type of fermion.
On the other hand, we have $f \ll 1$ for all known fermions,
except for the top quark where $f_t \sim 1$. This is in contrast with
the gauge couplings (at $W$ scale) $g_1 \cong 0.36$, $g_2 \cong 0.65$
and $g_3 \simeq 1.2$ for strong $SU(3)$.
It was suggested a long time ago \cite{radm} that
the observed fermion mass hierarchy
may be due to radiative mechanisms.
It is desirable to have Yukawa couplings $f \sim 1$, just like
gauge couplings.
Given that $\sqrt{2} m_\tau/v \cong 0.01$ is itself rather small,
we would naturally need {\it new} leptons to provide the ``seed"
for mass generation in the lepton sector.
Neutrino counting $N_\nu= 2.99 \pm 0.04$ and
direct search limits $m_E$, $m_N > M_Z/2$ \cite{PDG} imply that
new sequential leptons $E$ and $N$ are at scale $v$,
deviating sharply from earlier patterns \cite{3p1}.
In this Letter, we construct a simple model where
$E$ and $N$ receive mass at tree level, but all lower generation
lepton masses are generated by loop processes.
%including the possibility of Majorana mass for neutrinos.

We start with the lepton sector of minimal ``$3+1$" generations \cite{3p1},
where there is only one right-handed neutral lepton $N_R$.
Consider a discrete $Z_8$ symmetry ($\omega^8 = 1$).
We assign both $\bar\ell_{iL} = (\bar\nu_{iL}$, $\bar e_{iL}$)
and $e_{iR}$ to transform as
$\omega^3$, $\omega^2$, $\omega^1$, $\omega^4$ for $i = 1-4$,
respectively, while $N_R$ transforms as $\omega^4$.
The scalar sector consists of three doublets,
$\Phi_0,\ \Phi_3$ and $\Phi_5$,
transforming as $1$, $\omega^3$ and $\omega^5$, respectively.
Thus, aside from $E\simeq e_4$ and $N$,
only nearest-neighbor Yukawa couplings are allowed,
\begin{eqnarray}
  -\cal{L}_{\rm Y} &=&  f_{44} \bar\ell_{4L} e_{4R} \Phi_0
        +  \tilde f_{44} \bar\ell_{4L} N_{R}  \tilde\Phi_0 \nonumber \\
       &+& f_{43} \bar\ell_{4L} e_{3R} \Phi_3
        +  f_{34} \bar\ell_{3L} e_{4R} \Phi_3
        +  \tilde f_{34} \bar\ell_{3L} N_R \tilde\Phi_5   \nonumber \\
       &+& f_{32} \bar\ell_{3L} e_{2R} \Phi_5
        +  f_{23} \bar\ell_{2L} e_{3R} \Phi_5             \nonumber \\
       &+& f_{21} \bar\ell_{2L} e_{1R} \Phi_3
        +  f_{12} \bar\ell_{1L} e_{2R} \Phi_3 \ \ \ \ + H.c.
\end{eqnarray}
where $\tilde\Phi \equiv i \sigma_2 \Phi^*=( \phi^{0*}, -\phi^-)$
as usual. We assume $CP$ invariance for simplicity.

If only $\langle\phi_0^0\rangle = v/\sqrt{2}$,
$E$ and $N$ become massive
and are naturally at $v$ scale if $f_{44}$, $\tilde f_{44}\sim 1$.
The lower generation leptons remain massless at this stage,
protected by the $Z_8$ symmetry.
To allow for radiative mass generation,
the $Z_8$ symmetry is {\it softly} broken down to $Z_2$
in the Higgs potential by $\Phi_3$-$\Phi_5$ mixing.
Explicitly,
\begin{eqnarray}
 V = \sum_i &        \mu^2_i & \Phi_i^\dagger \Phi_i
    +\sum_{i,j} \lambda_{ij} (\Phi_i^\dagger \Phi_i) (\Phi_j^\dagger \Phi_j)
    +\sum_{i \neq j} \eta_{ij}(\Phi_i^\dagger \Phi_j) (\Phi_j^\dagger \Phi_i)
                                                        \nonumber \\
   +\, [    & \tilde \mu^2 & \Phi^\dagger_3 \Phi_5
    + \zeta (\Phi_0^\dagger \Phi_3) (\Phi_0^\dagger \Phi_5)
                                     + H.c.].
\end{eqnarray}
Note that the $\zeta$ term is $Z_8$ invariant, while
the gauge invariant ``mass" $\tilde \mu^2$ transforms as $\omega^2$.
Since only $\mu_0^2 < 0$, while $\mu^2_3$ and $\mu^2_5 >0$,
$\phi^0_0 \rightarrow (v+H_0+i\chi_0)/\sqrt 2$, and $\phi^0_i \rightarrow
(h_i+i\chi_i)/\sqrt 2$ for $i=3,\ 5$. The quadratic part of
$V$ is
\begin{eqnarray}
  V^{(2)} &=& \lambda_{00} v^2 H^2_0
           + \sum_{i\neq 0} \left(
              {1\over 2}(\mu^2_i+\lambda_{0i}v^2+\eta_{0i}v^2)(h^2_i+\chi^2_i)
             +(\mu^2_i+\lambda_{0i}v^2)\vert\phi_i^+\vert^2 \right)
                                                             \nonumber \\
          &+& \tilde \mu^2 (h_3 h_5+\chi_3\chi_5
                           +\phi^-_3 \phi^+_5 + \phi^-_5 \phi^+_3)
           + {1\over 2} \zeta v^2(h_3 h_5-\chi_3\chi_5).
\end{eqnarray}
The standard Higgs boson $H_0$ couples only diagonally to heavy particles.
The nonstandard scalars $(\phi^\pm_3, \phi^\pm_5)$,
$(h_3, h_5)$ and $(\chi_3, \chi_5)$ mix via $\tilde\mu^2$ and $\zeta$ terms.
Rotating by
$\theta_+$, $\theta_H$ and $\theta_A$, we obtain the mass basis
$(H^+_1, H^+_2)$, $(H_{1}, H_{2})$ and $(A_{1}, A_{2})$, respectively.
It is clear that $\sin\theta_+ \rightarrow 0$, $(\theta_A, m_{A_1},
m_{A_2}) \rightarrow (-\theta_H, m_{H_1}, m_{H_2})$ as $\tilde\mu^2\to 0$,
while in the limit $\zeta \rightarrow 0$,
$(\theta_A, m_{A_1}, m_{A_2}) \rightarrow (+\theta_H, m_{H_1}, m_{H_2})$.
These two limits restore the two extra $U(1)$ symmetries of
the doublets $\Phi_3$ and $\Phi_5$.
%and would hamper radiative mass generation, as we shall see.
%$\phi^\pm_3$ and $\phi^\pm_5$ would not mix,
%while $H_3$, $H_5$ and $\chi_3$, $\chi_5$ would still
%pair up as two complex neutral scalars.
As we shall see, fermion mass generation is due to mixing
and nondegeneracy of the two charged scalars,
and especially the four real scalar fields.

The $\tau$ lepton acquires mass via the one-loop diagram
shown in Fig. 1,
\begin{eqnarray}
  m_{33} = \left( {{\tilde f_{34} f_{43}} \over {32 \pi^2}} \right)\,
                & & \ \ \sin 2\theta_+\,
                                [G(m_{H_1^+}/m_N) - G(m_{H_2^+}/m_N) ]\; m_N
                                                               \nonumber \\
         + \left( {{f_{34} f_{43}} \over {32 \pi^2}} \right)
                & & \ \left[ \cos^2\theta_H\;
                                 G(m_{H_1}/m_E) \right.
                        +  \sin^2\theta_H\,
                                 G(m_{H_2}/m_E)        \nonumber \\
                & &  -     \cos^2\theta_A\
                                 G(m_{A_1}/m_E) \left.
                        -  \sin^2\theta_A\,
                                 G(m_{A_2}/m_E) \right]\, m_E,
\end{eqnarray}
where $G(x)=(x^2\ln x^2)/(x^2-1)$,
while $m_{34} = m_{43} = 0$.
As a numerical exercise, let $f_{43} = \tilde f_{34} =
f_{44} = \tilde f_{44} \sim \sqrt{2}$
so $\tilde f_{34} f_{43}/4\pi \simeq 1/2\pi$ and $m_N,\ m_E \cong 246$ GeV.
Then $\sin 2\theta_+\, (G(x_1)-G(x_2)) \simeq 0.6$ would make
$m_\tau=m_{33}\simeq 1.8$ GeV,
if the $m_E$ term contributes as much as the $m_N$ term,
which is usually the case.
There are separate GIM-like cancellation mechanisms rooted
in charged and neutral scalar mixing.
%These are remnants of the two extra $U(1)$ symmetries that would
%emerge in the limit of vanishing $\tilde \mu^2$ or $\zeta$.
In general $G(x_1) - G(x_2)$ is regulated by $\sin 2\theta_+$, hence
typically $\sin 2\theta_+\, (G(x_1)-G(x_2)) \ltap 1$.
Similar statements can be made for the neutral scalar contribution.
This implies that $f_{43},\ \tilde f_{34},\ f_{44}, \tilde f_{44}
\sim f \gtap 1$ is {\it natural} in our model.

If $\tilde\mu^2\rightarrow 0$, both terms would vanish and $m_\tau = 0$.
It is interesting to note that even if $\tilde \mu^2 \neq 0$,
the neutral scalar contribution would vanish if $\zeta = 0$.
This is of crucial importance for muon and electron mass generation,
for they receive radiative masses at
two- and three-loop order, respectively,
via neutral scalar loops that are similar to Fig. 1(b).
These ``nested" diagrams are illustrated in Fig. 2.
The upshot then is that $\zeta v^2/2$ should be of similar order of
magnitude as $\tilde \mu^2$, which in turn implies that
$\mu_3^2$ and $\mu_5^2$ should also be of order $v^2$.
Hence, nonstandard Higgs boson masses cannot be too
far above the electroweak scale!

Since off-diagonal mass terms $m_{24}$ and $m_{13}$
are also at two- and three-loop order, respectively,
we have the mass hierarchy
$m_E : m_\tau : m_\mu : m_e \sim 1 : \lambda : \lambda^2 : \lambda^3$,
where $\lambda$ is the loop expansion parameter.
That is, $m_i/m_{i+1} \sim f_{i,i+1}f_{i+1,i}/{32\pi^2}$ or more.
If $\sqrt {f_{i,i+1}f_{i+1,i}} \sim 1$ for all $i = 1-3$,
the mass hierarchy of order $10^{-1}-10^{-2}$ can be realized.
Together with $f_{44}$, $\tilde f_{44} \sim 1$,
we see that Yukuwa couplings could be
generation blind.
It is amusing that in our model, {\it all} dimensionless couplings
seem to be of order one, and {\it all} scale parameters are
of order $v$.

It is intriguing that the model could account for
$m_e/m_\mu\ (\sim m_\tau/m_E) \sim 1/200 \ll m_\mu/m_\tau
\sim 1/20$. Concentrating on neutral scalar loops,
from Figs. 1 and 2 we see that $m_e$, $m_\tau$ come
from $m_\mu$, $m_E$ seeds via ``$\phi_3^0$" loop,
while $m_\mu$ arises from $m_\tau$ seed via ``$\phi_5^0$" loop.
%In mass basis this means the interchange
%$\cos^2\theta_{H,A} \leftrightarrow \sin^2\theta_{H,A}$.
With obvious notation, we estimate
$m_e/m_\mu \simeq \lambda\; \left[s_H^2\ln(m_{H_2}^2/m_{H_1}^2)
- s_A^2\ln(m_{A_2}^2/m_{A_1}^2) + \ln(m_{H_1}^2/m_{A_1}^2)\right]$,
while for $m_\mu/m_\tau$ one has $s_{H,A}^2 \leftrightarrow c_{H,A}^2$.
As an example, we could have $m_{H_1} \sim m_{A_1} \sim m_{A_2}$,
then $(m_\mu/m_\tau)/(m_e/m_\mu) \cong \cot^2\theta_H \simeq 12$,
and $\sin\theta_H \simeq 0.28$ which is rather reasonable.
%Thus, $m_\mu/m_\tau$ being larger than $m_e/m_\mu$ (or $m_\tau/m_E$)
%comes about from having two distinct extra Higgs doublets in the model.

Note that $N_R$ is introduced solely for the purpose of satisfying
LEP bound \cite{3p1}. Having just a massive $E$ would have
been sufficient for charged lepton mass generation.
However, $N_R$
%is a gauge singlet and transforms as $\omega^4$
%under $Z_8$, it
could in principle have Majorana mass $m_R$,
which could serve as seed for radiatively generating Majorana mass
for the three left-handed neutrinos.
We emphasize that $m_R \gg m_N$ is not allowed \cite{3p1},
for then the seesaw mechanism \cite{seesaw} would
drive $N_L$ mass effectively to zero, violating LEP bound.
Rough estimates of loop induced Majorana neutrino masses
indicate that $m_R$ should be rather small, and we set $m_R = 0$
in the present work.
However, our model provides interesting, new mechanisms,
details of which will be reported elsewhere.

We turn to phenomenological prospects. These depend on the
lowest allowed mass(es) for the charged or FCNC neutral Higgs bosons.
Mixing effects in the charged current \cite{3p1}can be ignored
since they are radiatively generated.
%Na\"\i vely one might have expected arbitrary
%charged scalar and flavor changing neutral (FCNC) scalar boson
%masses since $\mu_3^2$, $\mu_5^2 > 0$ appear as free parameters.
%However,
As discussed earlier, radiative mass generation
for $\mu$ and $e$ suggest that
nonstandard Higgs boson masses should not be too far
above the electroweak scale $v$. Hence, one might worry about
low energy FCNC effects.
Very stringent limits exist for $\mu\to e\gamma$ \cite{PDG}.
Interestingly, our model has a dichotomy \cite{Z4} of leptons:
$E$, $N$, $\mu$, and $\nu_\mu$ are even under $Z_2$,
whereas $\tau$, $e$, $\nu_\tau$, and $\nu_e$ are odd.
For scalars, $\Phi_0$ is even, while $\Phi_3$ and $\Phi_5$ are odd.
Hence, $\mu \to e\gamma$ is {\it forbidden} in our model,
since the photon is $Z_2$ invariant.
Similarly, $\tau \to \mu\gamma$ is forbidden,
but $\tau \to e\gamma$ is allowed, as we shall discuss later.

The present experimental errors \cite{PDG} on $g-2$ for $e$ and $\mu$
imply \cite{MNW} lower bounds of a few
hundred GeV on the {\it effective} mass of the exchanged scalar bosons.
Standard $\mu$ and $\tau$ decay modes such as $\mu\to e\nu\nu$ and
$\tau\to\mu\nu\nu$ do not give better constraints
because of uncertainty in $M_W$ measurement \cite{Z4}.
The most stringent constraint on our model
turns out to be from leptonic FCNC $\tau$ decays:
$\tau \rightarrow \mu\mu^\mp e^\pm$
(the decay modes $\tau \rightarrow e e^\pm \mu^\mp$ are
forbidden).
Each has four contributions, two of which are shown in Fig. 3
for $\tau^- \rightarrow \mu^- \mu^- e^+$.
The inverse effective mass squared
$1/m^2(\phi_{3,5}^0)$ corresponds to a sum over
neutral (pseudo)scalars $\sum_{i=1}^4 a_i/m_i^2$ where
we now order according to mass $m_i$,
while $a_i$ are mixing factors that should satisfy
$0 < \vert a_i\vert \leq 1$, $\sum_i a_i = 0$.
They are nothing but $\pm s_{H,A}\, c_{H,A}$.
Thus, in general $m^2(\phi_{3,5}^0) > m_1^2$, the lightest (pseudo)scalar
mass, and could be much larger than $v^2$.
Assuming single (lightest) channel dominance, we find
\begin{equation}
B(\tau^-\rightarrow \mu^-\mu^- e^+) \simeq
   {1\over 2} \left( f_{e\mu}f_{\mu\tau}\, a_1/m_1^2
             \over g^2/M_W^2 \right)^2\, B(\tau\to \mu\nu\bar\nu),
\end{equation}
where
$2 f^2_{e\mu}f^2_{\mu\tau} \equiv \{(f_{23}f_{12})^2 + (f_{32}f_{21})^2\}$.
We show in Fig. 4 $B(\tau^- \to \mu^- \mu^- e^+)$
{\it vs.} $\sqrt{f_{e\mu}f_{\tau\mu}\, \vert a_1\vert}$ for
$m_1 = (0.5,\ 1,\ 2,\ 4)\, v$ ($\cong $ 125, 250, 500, 1000 GeV).
The present experimental bound of
$B(\tau^- \rightarrow \mu^- \mu^- e^+) < 1.6 \times 10^{-5}$
 \cite{PDG} is also shown. We see that, because of $\vert a_1\vert < 1$
and cancellation effects,
it is possible to have all Yukawa couplings of order unity
while physical nonstandard scalar masses are of order $v$ or greater.
Our model could naturally account for $B(\tau^-\rightarrow \mu^-\mu^- e^+)$
just below $10^{-5}$.
%To be more specific, the bound implies
%$\{(f_{23}f_{12})^2 + (f_{32}f_{21})^2\}^{1/4}
%                     \ltap
%       [m(\phi_{3,5}^0)/{1\ {\rm TeV}}]$.
Similar results are obtained
%for $\{(f_{23}f_{21})^2  +(f_{32}f_{12})^2\}^{1/4}$
from $B(\tau^- \rightarrow \mu^- \mu^+ e^-) < 2.7 \times 10^{-5}$ \cite{PDG}.
These decays are exceptionally clean, and should be searched for vigorously.

The $\tau^- \to \mu^-\mu^+e^-$ decay leads to $\tau\to e\gamma$ at
one-loop order.
We find
\begin{equation}
{B(\tau\to e\gamma)\over B(\tau\to\mu\bar\mu e)}
\ltap 24{\alpha\over \pi}\, \left({m_{\mu}\over m_{\tau}}\right)^2\,
\left(\ln {m_{\mu}^2\over m^2_1} \right)^2,
\end{equation}
where we assume same channel dominance, and drop
a constant term accompanying the large logarithm.
As an estimate, the ratio is less than $1/21 - 1/15$ for
$m_1 \simeq 250$ GeV $- 1$ TeV,
%The large logarithm compensates
%$\cal{O}(\alpha)$ and chirality flip ($m_{\mu}^2/m_{\tau}^2$) suppression,
hence $\tau\to e\gamma$ is typically just one order of magnitude below
$\tau\to\mu\bar\mu e$, {\it i.e.} at $10^{-6}$ or lower.
The present experimental bound is $\sim 10^{-4}$ \cite{PDG}.

In our model neutral scalars couple to
$\mu\bar e$ and $\bar\mu e$ simultaneously, therefore they
mediate muonium-antimuonium conversion \cite{HM}
(without doubly charged Higgs!).
Unlike $\tau\to \mu\mu^\mp e^\pm$ which is mediated by
$\phi_3^0$-${\phi_5^0}^{\left(*\right)}$ mixing,
here we have a $\phi_3^0$ mediated process.
The scalar mixing factors $a_i$ are of same sign
{\it i.e.} of the form $+c_{H,A}^2$, $+s_{H,A}^2$ and $\sum_i a_i = 2$.
Assuming that the dominating scalar has mass $\sim v$ and
the mixing factor ranges from $0.01 - 1$,
we estimate that the effective four Fermi coupling
is of order $(0.001 - 0.1)\, G_F$, compared with
the present bound of $0.16\, G_F$ \cite{PDG}.
The limit  may be improved to  $10^{-3}\, G_F$ soon \cite{PSI}.
Note that we have an unusual effective interaction
$\bar\mu (1-\gamma_5) e\, \bar\mu(1+\gamma_5) e$.
In the same vein, the process $\mu\to e\nu_e\bar\nu_\mu$
is mediated by $\phi_3^+$, and has a
four Fermi coupling of similar order.
The present bound is $0.14\, G_F$, but may be pushed down to
$10^{-2}\, G_F$ \cite{lampf} in near future.

Consider now the decays of the heavy lepton E.
If $m_E < m_N$, since charged current mixing is loop suppressed,
$\phi_{3,5}^\pm$-induced
$E \to \nu_\tau (e \bar\nu_\mu,\ \mu \bar\nu_e,\
\mu \bar\nu_\tau,\ \tau \bar\nu_\mu)$
and $\phi_{3,5}^0$-induced FCNC
$E \to \tau (e^\pm \mu^\mp, \mu^\pm \tau^\mp)$
decays could be dominant.
They could still be prominent %($10^{-2}$ or greater)
for $m_E \gtap m_N$ since $W$-induced transitions such as
$E\rightarrow N e \bar\nu_e$ are kinematically suppressed
until $m_E - m_N$ approaches $M_W$.
However, for $m_E\gtap m_N$, new scalar induced decays
such as $N\to \nu_\tau \mu^\pm e^\mp$ would dominate $N$ decay.
Since the lightest scalar might be lighter than $m_E$ or $m_N$,
it may even be produced directly in $E$, $N$ decay.
These decays would provide dramatic signatures at future colliders.

In summary, we have presented a realistic model for
radiative lepton mass generation.
The model has $Z_8$ symmetry and three Higgs doublets
with nearest-neighbor Yukawa couplings. Only
$E$ and $N$ receive tree level masses upon spontaneous symmetry breaking.
When $Z_8$ is softly broken down to $Z_2$, they provide
the seed for generating, {\it order by order},
loop masses for $\tau$, $\mu$ and $e$.
Yukawa couplings are naturally of order unity, and
the mass pattern
$m_E : m_\tau : m_\mu : m_e \simeq 1 : \lambda : \lambda^2 : \lambda^3$
emerges, which suggests the {\it possibility of
universal Yukawa couplings}.
The model could account for $m_\mu/m_\tau \gg m_e/m_\mu,\ m_\tau/m_E$
as a consequence of mixing effects among nonstandard Higgs bosons.
The residual $Z_2$ symmetry forbids
$\mu\to e\gamma$ and $\tau\to \mu\gamma$ type of transitions.
The charged and FCNC neutral Higgs bosons have weak scale masses.
However, due to GIM-like cancellations among themselves,
they could mimic TeV scale physics.
The most promising channels seem to be FCNC tau
decays $\tau \rightarrow \mu \mu^\pm e^\mp$
(but not $\tau \rightarrow e e^\pm \mu^\mp$)
and $\tau\to e\gamma$ at just below $10^{-5}$ and $10^{-6}$,
respectively.
Muonium-antimuonium conversion and $\mu\to e\nu_e\bar\nu_\mu$
could have strength $(0.001 - 0.1)\, G_F$.
FCNC leptonic decays of the fourth generation $E$ and $N$
are likely to be quite prominent.
We have consistently assumed that the Majorana mass $m_R= 0$
for $N_R$. A small $m_R$ could induce Majorana masses for left-handed
neutrinos via loop processes.
%The $Z_2$ symmetry may permit $\nu_e-\nu_\tau$ oscillations in this model.
The quark sector is clearly richer but more difficult.
Work is in progress, and will be reported elsewhere.

\acknowledgments
We thank E. Ma, D. Chang and C.-Q. Geng for useful discussions.
The work of GGW is supported in part by grant NSC 83-0208-M-002-025-Y,
and WSH by NSC 82-0208-M-002-151
of the Republic of China.

\vskip -1cm
\figure{Mechanism for $m_\tau$.}
\vskip -1cm
\figure{Mechanisms for $m_\mu$ and $m_e$ via ``nested" neutral scalar
        diagrams.}
\vskip -1cm
\figure{Some diagrams contributing to $\tau \rightarrow \mu \mu^- e^+$.}
\vskip -1cm
\figure{$B(\tau^- \rightarrow \mu^-\mu^- e^+)$ {\it vs.}
        $\sqrt{f_{e\mu}f_{\tau\mu}\, \vert a_1\vert}$ for
        (top to bottom) $m_1 =$ 125, 250, 500, 1000 GeV.
        Straight line indicates present limit.}
%\vskip 2cm \LARGE
%\mbox{\boldmath $B(\tau^- \rightarrow \mu^-\mu^- e^+)$}\\
%\vskip 1cm \hskip 1cm
%\mbox{\boldmath $\sqrt{f_{e\mu}f_{\mu\tau}\, \vert a_1\vert}$}
\end{document}